# SPATIAL AND FREQUENCY DEPENDENCIES OF LOCAL PHOTORESPONSE OF HTS STRIP-LINE RESONATOR IN THE REGIME OF TWO-TONE MICROWAVE INTERMODULATION EXCITATION

Alexander P. Zhuravel

B. Verkin Institute for Low temperature Physics & Engineering, NAS of Ukraine

Address: 47 Lenin Avenue, Kharkov, 61103, Ukraine

Tel.: +38(057)341-0907, Fax: +38(057)340-3370, E-mail: zhuravel@ilt.kharkov.ua

Steven M. Anlage

Center for Nanophysics and Advanced Materials, Department of Physics, University of Maryland

Address: College Park, MD 20742-4111 USA

Tel.: (301) 405-7321, Fax: (301) 405-3779, E-mail:anlage@umd.edu

Alexey V. Ustinov

Physikalisches Institut, Karlsruhe Institute of Technology (KIT)

Address: Wolfgang-Gaede-Str. 1, D-76131, Karlsruhe, Germany

Tel.: +49-721-608-3455, Fax: +49-721-608-6103, E-mail: ustinov@physik.uni-karlsruhe.de

High-$T_c$ superconducting (HTS) films have been widely applied in passive microwave electronics including high-Q resonators for mobile telephony. One of the crucial problems in applying such HTS devices remains their nonlinear (NL) response with respect to pumping power. This manifests itself as *global* intermodulation product distortion (IMD) when two or more high-frequency (HF) signals are NL mixed in the resonant frequency band. A predominately *local* origin of NL sources through *global* NL media motivates the development of spatially resolved methods to identify the dominant mechanisms of NL response. We have previously shown [1] that by using different imaging contrast modes of low-temperature (LT) laser scanning microscopy (LSM) one can solve this problem with great success. Additionally, a new paradigm of HF frequency-dependent singularity of LSM images was developed to spatial describe (i) local changes in surface resistance $\delta R_s(x, y)$ and (ii) HF current density distribution $J_{HF}(x, y)$ as a function of resonator $(x, y)$ surface coordinates by using a method of spatially-resolved complex impedance partition [2]. Here, we propose a new phenomenological approach to spatially-resolved research of NL microwave properties of operating thin-film superconducting resonators. The approach is based on frequency and spatial uniqueness of LSM images that can be extracted from a set of 2-D patterns representing $x$-$y$ distribution of the LSM photoresponse, $PR_f(x, y)$, at fixed third-order IMD frequencies $2f_1$-$f_2$ and $2f_2$-$f_1$ as a result of two-tone resonator microwave excitation at equidistant frequencies $f_1$ and $f_2$ relative to the fundamental resonance, $f_0$. We show that similar to what is described in [2] manipulation of LSM images at $2f_1$-$f_2$ and $2f_2$-$f_1$ can be used to present NL components of IMD LSM $PR(x, y)$ in terms of its independent spatial variations of (i) inductive $PR_{IMD,X}(x, y)$ and (ii) resistive $PR_{IMD,R}(x, y)$ contributions. In particular the realistic spatial coherence between such IMD components and $J_{HF}(x, y)$ flow, so crucial for determining the NL properties of the device, have not been investigated.

The tested device was a half wave $YBa_2Cu_3O_{7-\delta}$ (YBCO) microstrip (with thickness $d = 1$ μm) resonator on a 500 μm $LaAlO_3$ ($\varepsilon_r$=24.2) substrate (LAO) designed for operation at $f_0$=1.85 GHz with $Q$~1230 at T=78 K. The resonator was connected to the IMD scheme of a two-tone excitation centered with 1 MHz spacing around $f_0$. For LSM imaging, the device was $x$-$y$ scanned by a 4 μm diameter thermal probe produced by a 1.1 μm diameter focused laser beam that is amplitude modulated with a frequency of about 100 kHz on the sample surface. Position of the LSM raster was centered not far from the peak of the standing wave current pattern near the YBCO strip edge to investigate an area with maximum $J_{HF}(x, y)$ gradient that is the most suitable for the following spatially-resolved analysis. All LSM images were acquired at $T$=78 K and at $P_{f1}$=$P_{f2}$=9 dBm using step-by-step scanning of the sample by the laser probe with equal distance between steps of about 0.4 μm along both $x$ and $y$ direction. The linear components of HF LSM $PR_f(x, y)$ were measures at $f_1$ and $f_2$ to make spatial separation between inductive $PR_X(x, y)$ and dissipative $PR_R(x, y)$ contribution as explained by [2]. The derived $PR_X(x, y)$ was used to present LSM image of $J_{HF}(x, y)$ calibrated in real units by using a procedure described in [1]. Changes in $\delta R_s(x,y)$ were estimated as $PR_R(x,y)/PR_X(x,y)$ considering that $PR_R(x, y) \sim J^2_{HF}(x, y) * \delta R_s(x, y)$ and $PR_R(x, y) \sim J^2_{HF}(x, y)$.

Additionally, two LSM $PR_{2f1-f2}(x, y)$ and $PR_{2f2-f1}(x, y)$ images were measured in the same area of the resonator at combinational frequencies $2f_1$-$f_2$ and $2f_2$-$f_1$ of IMD harmonics. The spectral dependence of IMD photoresponse was estimated by using an expression for global IMD power

$P_{IMD}=Q^4(P^2_{f1}P_{f2})(1/P^2_c)[1+(\Delta R/Q\Delta L)^2]$ that has been developed in [3]. Here, $P_c=2\pi fL_0I^2_{IMD}/\Delta L$ is the characteristic power, $L_0$ and $R_0$ are the linear parts, $\Delta L$ and $\Delta R$ are the nonlinearity coefficients in changes of the inductance $L=L_0+\Delta L(I/I_{IMD})^2$ and resistance $R=R_0+\Delta R(I/I_{IMD})^2$ per unit length of the microstrip related to the case when $j_{IMD}$ is on order of the critical current density $J_c$. We considered the frequency dependence of thermally induced modulation in $P_{IMD}$ in terms of the temperature derivative $\delta P_{IMD}(f)= (dP_{IMD}/dT)\delta T$, where separate contributions from $PR_{IMD,X}(x, y)$ and $PR_{IMD,R}(x, y)$ can be deduced through partial derivatives related to $f_0$ and $1/Q$, correspondingly:

$$PR_{IMD} \propto \frac{\partial P_{IMD}}{\partial T} = \frac{\partial P_{IMD}}{\partial(1/Q)}\frac{\partial(1/Q)}{\partial(T)} + \frac{\partial P_{IMD}}{\partial(f_0)}\frac{\partial(f_0)}{\partial(T)} \quad (1)$$

One can make sure that at frequencies very close to the fundamental resonance ($f \sim f_0$) dissipative and inductive components of LSM PR may be presented as:

$$PR_{IMD,R} = \frac{\partial P_{IMD}}{\partial(1/Q)} \propto \frac{-(1/Q)^3}{\left[4(f/f_0-1)^2+(1/Q)^2\right]^4} \quad (2a)$$

$$PR_{IMD,X} = \frac{\partial P_{IMD}}{\partial(f)} \propto \frac{(1/Q)^2}{f_0}\frac{(f/f_0-1)}{\left[4(f/f_0-1)^2+(1/Q)^2\right]^4} \quad (2b)$$

As evident from Eq. (2a), the spectrum of $PR_{IMD,R}(f)$ is approximated well by a symmetric (relative to $f_0$) Lorentzian-like curve of the same (negative) sign having equal magnitudes $PR_{IMD,R}(2f_1-f_2)= PR_{IMD,R}(2f_2-f_1)$ at equidistant IMD frequencies $2f_1-f_2$ and $2f_2-f_1$ around $f_0$. In contrast, the curve of $PR_{IMD,X}(f)$ dependence (see Eq. (2b)) crosses "0" at $f = f_0$ and has opposite signs i.e. $PR_{IMD,X}(2f_1-f_2)= (-)PR_{IMD,X}(2f_2-f_1)$ there. By virtue such an $f$-singularity of $PR_{IMD}(f)$, one can present LSM images of NL media as independent distributions of inductive and resistive NL sources by using the following transformations:

$$PR_{IMD,X}(x,y) = \left|\frac{PR_{IMD}(2f_1-f_2)-PR_{IMD}(2f_2-f_1)}{2}\right| \quad (3a)$$

$$PR_{IMD,R}(x,y) = \left|\frac{PR_{IMD}(2f_1-f_2)+PR_{IMD}(2f_2-f_1)}{2}\right| \quad (3b)$$

For the experimental verification of these statements, the same 40x20 μm² area of the resonator was visualized in different imaging modes of LSM contrast. In Fig. 1.

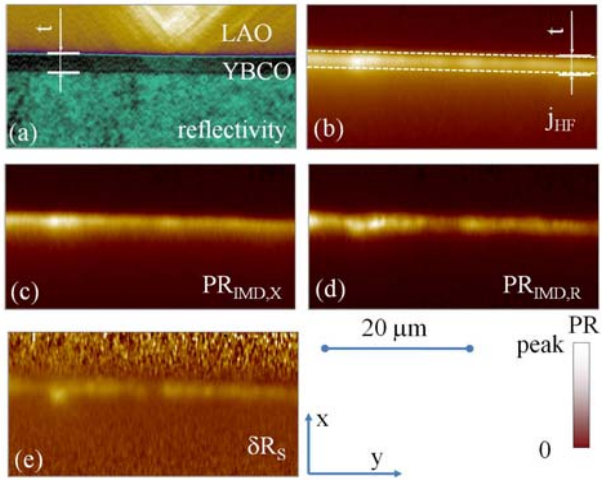

Fig.1. Set of LSM images that were analyzed and discussed in the text.

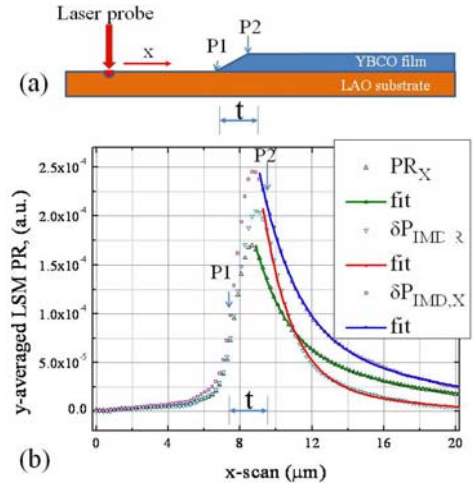

Fig.2. (a) Sketch of the sample cross section along x-scan and (b) profiles of LSM response there.

In Fig. 1(a) we have pictured an optical reflectivity LSM (false-color) map of the scanned area where the surface quality of the YBCO film as well as the twinned structure of the LAO substrate is clearly visible. The YBCO strip edge has a chamfered profile producing a darker YBCO reflectivity image within a t = 4 μm strip. Top (P1 in Fig. 2(a)) and bottom (P2 in Fig. 2(a)) edges of the chamfer are placed along two dashed white lines in Fig. 2(b) where an LSM image of $J_{HF}(x, y)$ is shown. The brightest area there coincides with maximum amplitude of $J_{HF}(x, y) \sim 10^9 A/m^2$ while the darkest ones correspond to zero current density. Spatial coherence between (i) $J_{HF}(x, y)$, nonlinear (ii) inductive $PR_{IMD,X}(x, y)$ and (iii) resistive $PR_{IMD,R}(x, y)$ components of LSM PR as well as (iv) $\delta R_s(x,y)$ is evident as one can see through comparison of all images presented in Fig. 1(b-e).

Fig. 2(b) shows typical profiles of different IMD and HF components of LSM PR(x, y = 4) together with their exponential fit that were measured along a single x-scan (see dashed line numbered as '4' in Fig. 4(a)) at fixed y(=4 lines)-cross-section. Similar profiles were used in Fig. 3 for log-log plotting of both local $PR_{IMD,R}$ and $PR_{IMD,X}$ responses as a function of local circulating HF power that can be approximated by the linear term of inductive LSM PR. As seen all four cross-sections (4, 18, 32 and 100 from Fig. 4(a)) of the resonator demonstrate a similar tendency for a linear increasing of $PR_{IMD,X}$ and square-low growth of $PR_{IMD,R}$ with respect to $PR_{HF,X}$ that is linked directly to IMD components through the same position of the laser probe. Also, the dominant contribution of $PR_{IMD,X}$ to the total IMD PR is seen at small $J_{HF}(x, y)$ closer to YBCO center. However at peak values of HF current density near the HTS film edge those amplitudes of NL response are almost equal. We believe that IMD LSM PR originates from a small laser-probe induced modulation $\delta P_{IMD}(x, y)$. of intermodulation power. In this case, $PR_{IMD,X}(x, y)$ and $PR_{IMD,R}(x, y)$ components show local NL properties of the microwave device in terms of corresponding local derivatives $\delta P_{IMD,X}(x, y)$ and $\delta P_{IMD,R}(x, y)$. Spatial integration (see Fig. 4(b)) shows that below $J_c(x, y)$ (see, for example data in Fig. 3(c) below P1), the power of real (dissipative) NL sources follows a classical cubic dependence $P_{IMD,R} \sim P^3_{IN}$ on input HF power. There were no changes in local values of the surface resistance found at all $J_{RF}(x, y) < J_c(x, y)$. Surprisingly, local sources of imaginary (inductive) IMD response show almost a square-law power dependence (see Fig. 4(b)) even below $J_c(x, y)$ except very close to $J_{RF}(x, y) \sim 0$. This contradicts previously published data considering $J_c(x, y)$ as the correct scale for the lower limit of NLs.

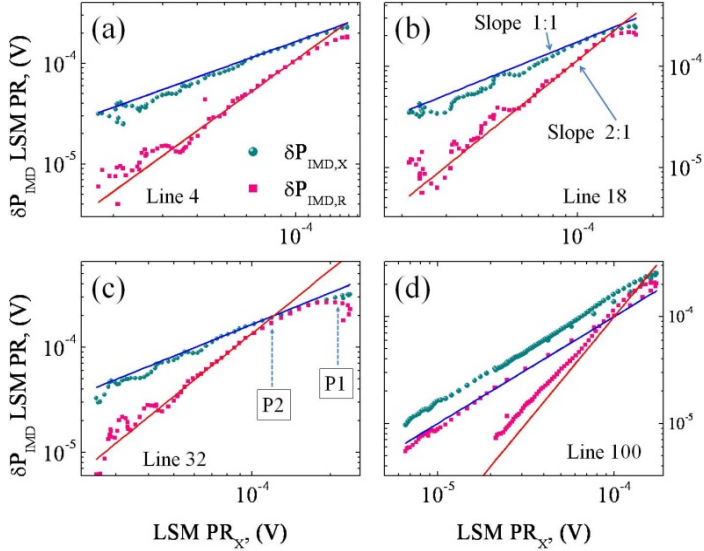 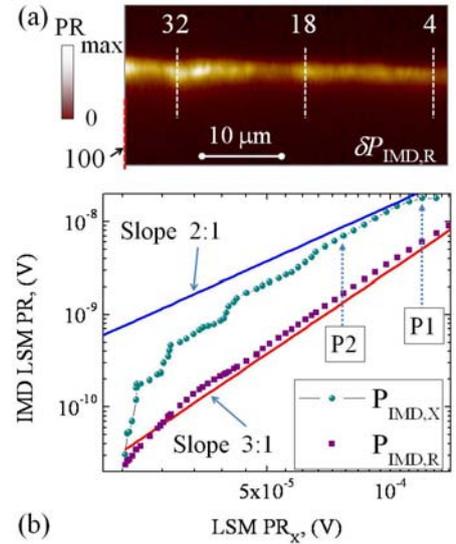

Fig.3. Log-log plots of IMD components of LSM photoresponse as a function of linear inductive one. The line cuts in parts (a) – (d) correspond to those shown in Fig. 4.

Fig.4. (a) LSM image of $PR_{IMD,R}(x, y)$ showing line-scans analyzed in Fig. 3 and (b) restored $P_{IMD}(P)$ dependencies.

The authors acknowledge the support of the Fundamental Researches State Fund of Ukraine and German International Bureau of the Federal Ministry of Education and Research (BMBF) under grant project UKR08/011 and a NASU program on "nanostructures, materials and technologies".